\documentclass[12pt]{article}%
\usepackage{graphics}%
\begin{document}%
\def\e{{\mathrm e}}%
\def\g{{\mbox{\sl g}}}%
\def\Box{\nabla^2}%
\def\d{{\mathrm d}}%
\def\R{{\rm I\!R}}%
\def\ie{{\em i.e.\/}}%
\def\eg{{\em e.g.\/}}%
\def\etc{{\em etc.\/}}%
\def\etal{{\em et al.\/}}%

\markright{Quasi-particle creation by analogue black holes\hfil}%
\title{\bf \LARGE Quasi-particle creation by \\ analogue black holes}%
\author{Carlos Barcel\'o~$^*$, Stefano Liberati~$^\dagger$,
Sebastiano Sonego~$^\ddagger$, and Matt Visser~$^\S$%
\\[2mm]%
{\small\it%
\thanks{\tt carlos@iaa.es}%
\ Instituto de Astrof\'{\i}sica de Andaluc\'{\i}a, CSIC,
Camino Bajo de Hu\'etor 50, 18008 Granada, Spain}%
\\[4mm]%
{\small\it%
\thanks{\tt liberati@sissa.it; http://www.sissa.it/\~{}liberati}%
\ International School for Advanced Studies,
Via Beirut 2-4, 34014 Trieste, Italy}
\\
{\small and}
\\
{\small\it INFN, Trieste, Italy}
\\[4mm]%
{\small\it%
\thanks{\tt sebastiano.sonego@uniud.it}%
\ Universit\`a di Udine, Via delle Scienze 208, 33100 Udine, Italy}%
\\[4mm]%
{\small\it%
\thanks{\tt Matt.Visser@mcs.vuw.ac.nz; http://www.mcs.vuw.ac.nz/\~{}visser}
\ School of Mathematics, Statistics,  and Computer Science,
Victoria University of Wellington,}%
\\%
{\small\it%
 Wellington, New Zealand}}%
\date{{\small 11 April 2006; {\LaTeX-ed \today}; gr-qc/0604058}}%
\maketitle%
\begin{abstract}%

We discuss the issue of quasi-particle production by
``analogue black holes'' with  particular attention to the
possibility of reproducing Hawking radiation in a laboratory.  By
constructing simple geometric acoustic models, we obtain a somewhat
unexpected result: We show that in order to obtain a stationary and
Planckian emission of quasi-particles, it is \emph{not}
necessary to create an ergoregion in the acoustic spacetime
(corresponding to a supersonic regime in the flow).  It is
sufficient to set up a dynamically changing flow \emph{either}
eventually generating  an arbitrarily small sonic region $v=c$, but
without any ergoregion, \emph{or} even just asymptotically,
in laboratory time, approaching a sonic regime with sufficient
rapidity.%

\vspace*{5mm}%
\end{abstract}%

\clearpage

\section{Introduction}%
\label{sec:intro}%
\setcounter{equation}{0}%

It is by now well established that the physics associated with
classical and quantum fields in curved spacetimes can be reproduced,
within certain approximations, in a variety of different physical systems
--- the so-called ``analogue models of General Relativity
(GR)''~\cite{analogue-book,living-review}.  The simplest example of
such a system is provided by acoustic disturbances propagating in a
barotropic, irrotational and viscosity-free fluid.%

In the context of analogue models it is natural to separate the
kinematical aspects of GR from the dynamical ones.  In general, within
a sufficiently complex analogue model one can reproduce any pre-specified
spacetime --- and the kinematics of fields evolving on it ---
independently of whether or not it satisfies the classical (or
semiclassical) Einstein equations~\cite{analogue-bec}.  Indeed, to date
there are no analogue models whose effective geometry is determined by
Einstein equations.  In this sense we currently have both
analogue spacetimes and analogues of quantum field theory in curved
spacetimes, but (strictly speaking) no analogue model for GR
itself~\cite{analogues-of-and-for}.%

In order to reproduce a specific spacetime geometry within an analogue
model, one would have to take advantage of the specific equations
describing the latter (for example, for fluid models, the Euler
and continuity equations, together with an equation of state), plus
the possibility of manipulating the system by applying appropriate
external forces. In the analysis of this paper we will think of the
spacetime configuration as ``externally given'', assuming that it has
been set up as desired by external means --- any back-reaction on the
geometry is neglected as in principle we can counter-balance its
effects using the external forces. In the context of analogue
models this is not merely a hypothesis introduced solely for
theoretical simplicity, but rather a realistic situation that is in
principle quite achievable.%

Specifically, in this paper we analyze in simple terms the issue of
quantum quasi-particle creation by several externally
specified $(1+1)$-dimensional analogue geometries simulating the
formation of black hole-like configurations. (In a previous
companion paper~\cite{companion} we investigated the causal
structure of these, and other, spacetimes.) In this analysis we have
in mind, on the one hand, the possibility of setting up laboratory
experiments exhibiting Hawking-like radiation~\cite{hawking,bd} and,
on the other hand, the acquisition of new insights into the physics
of black hole evaporation in semiclassical gravity.  All the
discussion holds for a scalar field obeying the D'Alembert wave
equation in a curved spacetime.  This means that we are not (for
current purposes) considering the deviations from the phononic
dispersion relations that show up at high energies owing to the
atomic structure underlying any condensed matter system.  We shall
briefly comment on these modifications at the end of the paper. For
simplicity, throughout the paper we adopt a terminology based on
acoustics in moving fluids (we will use terms such as acoustic
spacetimes, sonic points, fluid velocity, etc.), but our results are
far more general and apply to many other analogue gravity models not
based on acoustics.  We summarise the main conclusions below.%

First of all, we recover the standard Hawking result when
considering fluid flows that generate a supersonic regime at
finite time.  (That is, we recover a stationary creation of quasi-particles with a  Planckian spectrum.)  We then analyze the
quasi-particle creation associated with other types of
configurations.  In particular, we shall discuss in detail a
``critical black hole'' --- a flow configuration that
presents an acoustic horizon without an associated supersonic
region.  {From} this analysis we want to highlight two key
results:%
\begin{itemize}%
\item%
The existence of a supersonic regime (sound velocity $c$ strictly
smaller than fluid velocity $v$) is not needed in order to reproduce
Hawking's stationary particle creation. We demonstrate this fact by
calculating the quantity of quasi-particle production in
an evolving geometry which generates only an isolated sonic point
($v=c$), but without a supersonic region, in a finite amount of
laboratory time.%
\item%
Moreover, in order to produce a Hawking-like effect it is not even
necessary to generate a sonic point at finite time. All
one needs is that a sonic point develops in the asymptotic future
(that is, for $t\to +\infty$) \emph{with sufficient rapidity} (we shall
explain in due course what we exactly mean by this).%
\end{itemize}%

{From} the point of view of the reproducibility of a Hawking-like
effect in a laboratory, the latter result is particularly
interesting. In general, the formation of a supersonic regime in a
fluid flow --- normally considered to be the crucial requirement to
produce Hawking emission --- is associated with various different
types of instability (Landau instability in superfluids, quantized
vortex formation in Bose--Einstein condensates, etc.) that could mask
the Hawking effect.  To reproduce a Hawking-like effect without
invoking a supersonic regime could alleviate this situation.%

{From} the point of view of GR, we believe that our result could
also have some relevance, as it suggests a possible alternative
scenario for the formation and semiclassical evaporation of black
hole-like objects.%

The plan of the paper is the following: In the next section we
introduce the various acoustic spacetimes on which we focus
our attention, spacetimes that describe the formation of acoustic black holes
of different types.  In section~\ref{sec:creation} we present
separately the specific calculations of redshift for sound rays that pass
asymptotically close to the event horizon of these black holes.  By invoking
standard techniques of quantum field theory in curved spacetime, one
can then immediately say when particle production with a Planckian
spectrum takes place.  Finally, in the last section of the paper we
summarise and discuss the results obtained.%

\section{Acoustic black holes}
\label{sec:spacetimes}%
\setcounter{equation}{0}%

Associated with the flow of a barotropic, viscosity-free fluid
along an infinitely long thin pipe, with density and velocity
fields constant on any cross section orthogonal to the pipe, there
is a (1+1)-dimensional {\em acoustic spacetime\/} $({\cal M},\g)$,
where the manifold $\cal M$ is diffeomorphic to $\R^2$.  Using the
laboratory time $t\in\R$ and physical distance $x\in\R$ along the pipe as
coordinates on $\cal M$, the {\em acoustic metric\/} on $\cal M$
can be written as%
\begin{equation}%
\g=\Omega^2\left[-\left(c^2-v^2\right)\d t^2
+2\,v\,\d t\,\d x+\d x^2\right]
=
\Omega^2\left[-c^2\,d t^2
+(\d x+v\,\d t)^2\right]
\;,%
\label{metric}%
\end{equation}%
where $c$ is the speed of sound, $v$ is the fluid velocity, and
$\Omega$ is an unspecified non-vanishing function~\cite{visser98}.
In general, all these quantities depend on the laboratory
coordinates $x$ and $t$. Here, we shall assume that $c$ is a
constant. Hence, it is the velocity $v(x,t)$ that contains all the
relevant information about the causal structure of the acoustic
spacetime $({\cal M},\g)$.  We direct the reader to the companion
paper~\cite{companion} for a detailed analysis of the causal
structure associated with a broad class of $(1+1)$-dimensional
acoustic geometries, both static and dynamic.%

\subsection{Apparent horizon}%
\label{subsec:sonic}%

The sonic points, where $v(t,x)=\pm c$, correspond to the so-called
acoustic apparent horizons --- apparent horizons for the Lorentzian
geometry defined on $\cal M$ by the metric (\ref{metric}). The fact
of having an underlying Minkowski structure associated with the
laboratory observer makes the definition of apparent horizons in
acoustic models less troublesome than in GR (see
\eg\ reference \cite{living-review}, pp.\ 15--16).%

Consider a monotonically non-decreasing function $\bar{v}(x)$
such that $\bar{v}(0)=-c$ and $\bar{v}(x)\to 0$ for $x\to +\infty$.
If one chooses $v(x,t)=\bar{v}(x)$ in (\ref{metric}), the
corresponding acoustic spacetime represents, for observers with
$x>0$, a static black hole with the horizon located at $x=0$ (in
this case apparent and event horizon coincide), a black hole region
for $x<0$, and a (right-sided) surface gravity%
\begin{equation}%
\kappa:=\lim_{x\to 0^+}\frac{\d \bar{v}(x)}{\d x}\;.%
\label{kappa}%
\end{equation}%
We can, moreover, distinguish three cases:%
\begin{itemize}%
\item $\kappa\neq 0$ and $\bar{v}(x)<-c$ for $x<0$: a non-extremal black
hole;%
\item $\kappa\neq 0$ and $\bar{v}(x)=-c$ for $x<0$: a ``critical''
black hole;%
\item $\kappa=0$ and $\bar{v}(x)=-c$ for $x<0$: an extremal black hole.%
\end{itemize}%
Now, taking the above $\bar{v}(x)$, let us consider $t$-dependent
velocity functions%
\begin{equation}%
v(x,t)=\left\{\begin{array}{lll}%
\bar{v}(\xi(t))&\mbox{if}& x\leq\xi(t)\;,%
\\ %
\bar{v}(x)&\mbox{if}& x\geq\xi(t)\;,%
\end{array}\right.%
\label{velocity}%
\end{equation}%
with $\xi$ a monotonically decreasing function of $t$, such
that $\displaystyle\lim_{t \to -\infty} \xi(t) = +\infty$ and
$\displaystyle\lim_{t \to -\infty} \dot \xi(t) = 0$.  (The first
condition serves to guarantee that spacetime is flat at early times,
whereas we impose the second one only for simplicity. All the
analysis in the paper could be performed without adopting this
assumption, leaving the physical results unchanged. However,
that would require more case-by-case splitting, only to cover new
situations without physical interest.) There are basically two
possibilities for $\xi$, according to whether the value $\xi=0$ is
attained for a finite laboratory time $t_{\rm H}$ or asymptotically
for an infinite future value of laboratory time.%

\begin{figure}[htbp]%
\vbox{ \hfil \scalebox{0.50}{ {\includegraphics{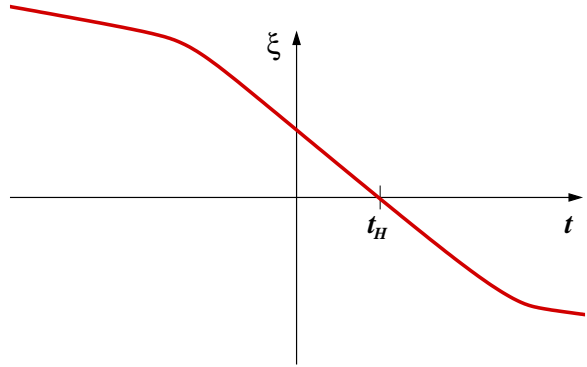}} }\hfil}%
\bigskip%
\caption{%
Plot of $\xi(t)$ for the formation of an acoustic apparent horizon at
a finite laboratory time $t_{\rm H}$. Only the behaviour of $\xi$ for
$t$ close to $t_{\rm H}$ is important.%
} \label{F:xi-bh}%
\end{figure}

In the first case $\xi(t_{\rm H})=0$ and the corresponding
metric (\ref{metric}) represents the formation of a non-extremal,
critical, or extremal black hole, respectively.  For small values
of $|t-t_{\rm H}|$ we have%
\begin{equation}%
\xi(t)=-\lambda\,(t-t_{\rm H})+{\cal O}([t-t_{\rm H}]^2)\;,%
\label{appxi-bh}%
\end{equation}%
where $\lambda$ is a positive parameter.  Hence the function
$\xi$ behaves, qualitatively, as shown in figure~\ref{F:xi-bh}.  %
Apart from this feature, the detailed behaviour of $\xi$ is largely
irrelevant for our purposes.%

\begin{figure}[htbp]%
\vbox{ \hfil \scalebox{0.50}{ {\includegraphics{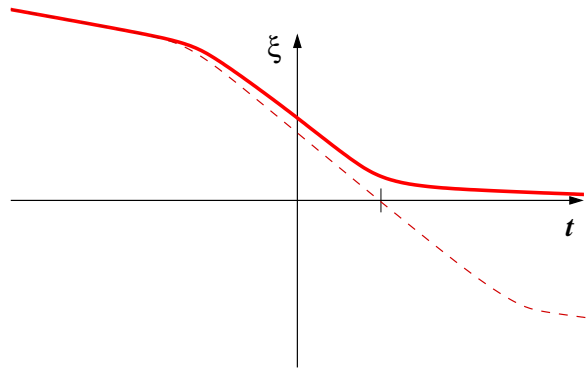}} }\hfil}%
\bigskip%
\caption{%
Plot of $\xi(t)$ for the asymptotic formation of an acoustic apparent
horizon at infinite future laboratory time. Only the asymptotic
behaviour of $\xi$ for $t\to +\infty$ is important.  (For comparison
we also plot $\xi(t)$, with a dashed line, for the formation of
the apparent horizon at finite laboratory time.)%
} \label{F:xi-cbh}%
\end{figure}

If instead $\xi\to 0$ is attained only at infinite future time, that
is $\displaystyle\lim_{t\to +\infty}\xi(t)=0$, one is describing the
asymptotic formation of either a critical black hole (if $\kappa\neq
0$; obviously, in this case choosing the non-extremal or the
critical $\bar v(x)$ profile is irrelevant) or an extremal black
hole (if $\kappa=0$).  Now the function $\xi$ behaves,
qualitatively, as shown in figure~\ref{F:xi-cbh}.  %
The relevant feature of $\xi(t)$ is its asymptotic behaviour as
$t\to+\infty$.  In the following we shall consider two
possibilities for this asymptotics, although others can, of course,
be envisaged:%
\begin{description}%
\item[\rm (i)] Exponential: $\xi(t)\sim A\,\e^{-\kappa_{\rm D}\,t}$,
with $\kappa_{\rm D}$ a positive constant, in general different from
$\kappa$, and $A>0$;%
\item[\rm (ii)] Power law: $\xi(t)\sim B\,t^{-\nu}$, with $\nu>0$
and $B>0$.%
\end{description}%
%

\subsection{Null coordinates}%
\label{subsec:redshift}%

For all the situations considered so far, spacetime is Minkowskian
in the two asymptotic regions corresponding to $t\to -\infty$, and
to $t\to +\infty$, $x\to +\infty$ (${\Im}^-$ and ${\Im}^+_{\rm
right}$, respectively, adopting the notation of
reference~\cite{companion}).  Starting with a quantum scalar field
in its natural Minkowskian vacuum at $t\to -\infty$, we want to know
the total quantity of quasi-particle production to be detected at
the right asymptotic region at late times, $t\to +\infty$, caused by the
dynamical evolution of the velocity profile $v(x,t)$.%

\begin{figure}[htbp]%
\vbox{ \hfil \scalebox{0.60}{ {\includegraphics{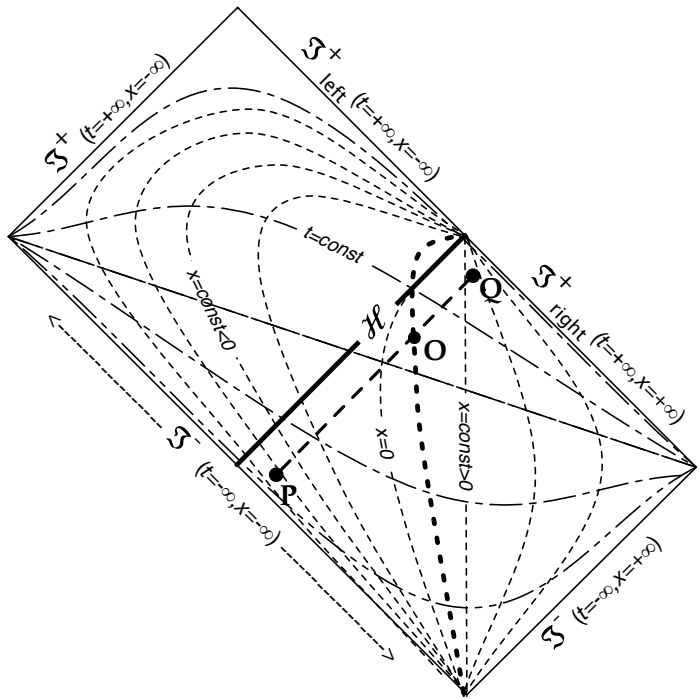}} }\hfil}%
\bigskip%
\caption{%
Conformal diagram of the spacetime corresponding to the formation of
an acoustic black hole.  The dotted line in bold represents the
worldline of the kink, $x=\xi(t)$.  The dashed straight line in bold
represents a right-going ray connecting the events $P=(x_i,t_i)$ near ${\Im}^{-}$ and $Q=(x_f,t_f)$ near ${\Im}^+_{\rm right}$.%
}\label{F:conf-bh}%
\end{figure}

In the geometric acoustic approximation, a right-going sound ray is
an integral curve of the differential equation%
\begin{equation}%
\frac{\d x}{\d t}=c+v(x,t)\;.%
\label{diffeq}%
\end{equation}%
We are interested in sound rays propagating from ${\Im}^{-}$ (see
figure~\ref{F:conf-bh}); that is, in solutions of (\ref{diffeq})
that satisfy an initial condition $x(t_i)=x_i$, with $x_i\sim c t_i$
in the limit $t_i\to -\infty$ (so $P:=(x_i,t_i)$ can be thought of
as an ``initial'' event corresponding to the emission of the
acoustic signal).  %
If such a ray ends up on ${\Im}^+_{\rm right}$, we can identify
``final'' events $Q:=(x_f,t_f)$ on it, with $x_f\sim c t_f$ as
$t_f\to +\infty$. For a ray connecting ${\Im}^-$ to ${\Im}^+_{\rm
right}$ one can also find an event $O:=(x_0,t_0)$ such that
$x_0=x(t_0)=\xi(t_0)$, which corresponds to the crossing of the
``kink'' in $v$, located at $x=\xi(t)$ according to equation
(\ref{velocity}), by the sound signal. Finally, we can define, for
such a ray, two parameters $U$ and $u$ as follows:%
\begin{equation}%
U:=\lim_{t_i\to -\infty}\left(t_i-x_i/c\right)\;;%
\label{U}%
\end{equation}%
\begin{equation}%
u:=\lim_{t_f\to +\infty}\left(t_f-x_f/c\right)\;.%
\label{u}%
\end{equation}%
Such parameters correspond to null coordinates in spacetime. If an
acoustic event horizon $\cal H$ is present in the spacetime, the
coordinate $U$ is regular on it (\ie, $U$ attains some finite value
on $\cal H$), whereas $u$ tends to $+\infty$ as $\mathcal H$ is
approached.%

We can express both $U$ and $u$ in terms of the velocity profile
(shape and dynamics) and of the crossing time $t_0$.  To this end, we can
integrate equation (\ref{diffeq}), first between $P$ and $O$:%
\begin{equation}%
\xi(t_0)-x_i=\int_{t_i}^{t_0}\d t\left(c+\bar{v}(\xi(t))\right)\;;%
\label{xi-x}%
\end{equation}%
then between $O$ and $Q$:%
\begin{equation}%
t_f-t_0=\int_{\xi(t_0)}^{x_f}\frac{\d x}{c+\bar{v}(x)}\;.%
\label{t-t}%
\end{equation}%
On replacing $x_i$ from equation (\ref{xi-x}) into (\ref{U}), we
find the value of $U$ for a generic right-moving ray that crosses
the kink at laboratory time $t_0$:%
\begin{equation}%
U=t_0-\frac{\xi(t_0)}{c}+\frac{1}{c}\int_{-\infty}^{t_0}\d
t\,\bar{v}(\xi(t))\;.%
\label{UU}%
\end{equation}%
Similarly, substituting $t_f$ from equation (\ref{t-t}) into
(\ref{u}), then adding and subtracting the quantity $\xi(t_0)/c$,
we find%
\begin{equation}%
u=t_0-\frac{\xi(t_0)}{c}-\frac{1}{c}\int_{\xi(t_0)}^{+\infty}\d
x\,\frac{\bar{v}(x)}{c+\bar{v}(x)}\;.%
\label{uu}%
\end{equation}%

In the analysis below, our chief goal consists of finding the relation between
$U$ and $u$ for a sound ray that is close to the horizon, \ie, in
the asymptotic regime $u\to +\infty$.  {From} such a relation it is
then a standard procedure to find the Bogoliubov $\beta$
coefficients and hence the total quasi-particle content to be
measured, in this case, by an asymptotic observer at ${\Im}^+_{\rm
right}$. (See, for example, reference~\cite{bd}). In the case of an
exponential relation between $U$ and $u$ it is a well established
result that a Planckian spectrum is observed at late
times~\cite{hu}, so Hawking-like radiation will be recovered.%

\section{Event horizon formation}%
\label{sec:horizon}%
\setcounter{equation}{0}%

When the apparent horizon forms at a finite laboratory time, say at
$t=t_{\rm H}$, an event horizon always exists, generated by the
right-moving ray that eventually remains frozen on the apparent
horizon, at $x=0$. For such a ray $t_0\to t_{\rm H}$, and since
$\xi(t_{\rm H})=0$, the $U$ parameter has the finite value%
\begin{equation}%
U_{\rm H}= t_{\rm H} +
\frac{1}{c}\int_{-\infty}^{t_{\rm H}}\d t\,\bar{v}(\xi(t))\;.%
\label{UH-fin}%
\end{equation}%
For a ray with $U<U_{\rm H}$ we then obtain, combining equations
(\ref{UU}) and
(\ref{UH-fin}):%
\begin{equation}%
U=U_{\rm H}+t_0-t_{\rm H}-\frac{\xi(t_0)}{c}
-\frac{1}{c}\int_{t_0}^{t_{\rm H}}\d t\,\bar{v}(\xi(t))\;.%
\label{U-fin}%
\end{equation}%
This \emph{exact} equation is now in a form suitable for conveniently
extracting \emph{approximate} results in the region $t_0\sim t_{\rm H}$,
corresponding to sound rays that ``skim'' the horizon.

On the other hand, when the trapping horizon consists of just one single sonic
point located at $t=+\infty$, it is not obvious that an event
horizon exists. Loosely speaking, in this case it might happen that
the trapping horizon form ``after'' every right-going ray from
${\Im}^-$ has managed to cross $x=0$.  Since there is a
competition between two infinite quantities --- the time at which
the trapping horizon forms, and the time at which the ``last''
right-going signal that connects $x=-\infty$ with $x=+\infty$
crosses $x=0$ --- a careful case-by-case analysis is in order.%

This is essentially all that can be said without relying on specific
features of $\bar{v}(x)$.  We now consider separately the various
situations of interest, focussing first on the issue of the existence
of the event horizon.%

\subsection{Non-extremal black hole}%
\label{subsec:bh-event}%

\begin{figure}[htbp]%
\vbox{ \hfil \scalebox{0.50}{ {\includegraphics{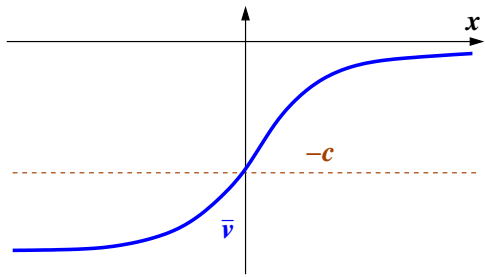}} }\hfil}%
\bigskip%
\caption{%
The static velocity profile $\bar{v}(x)$ for an acoustic
non-extremal black hole.%
} \label{F:v-bh}%
\end{figure}

\begin{figure}[htbp]%
\vbox{ \hfil \scalebox{0.50}{ {\includegraphics{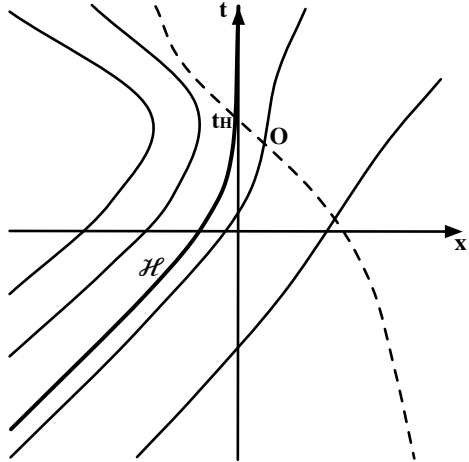}} }\hfil}%
\bigskip%
\caption{%
The worldlines of right-moving sound rays in the spacetime
describing the formation of a non-critical, non-extremal black
hole.  The thick solid line is the event horizon; the dashed
one is the worldline of the ``kink,'' $x=\xi(t)$.%
} \label{F:tx-bh}%
\end{figure}

In the case of a non-extremal black hole, the qualitative behaviour 
of the function $\bar{v}(x)$
is shown, graphically, in figure~\ref{F:v-bh}.  %
Note that, for small values of $|x|$, one can write%
\begin{equation}%
\bar{v}(x)=-c+\kappa\,x+{\cal O}(x^2)\;.%
\label{appv-bh}%
\end{equation}%
The function $\xi(t)$ behaves as already shown in
figure~\ref{F:xi-bh}. A sketch of the worldlines of right-moving
sound rays is presented in figure~\ref{F:tx-bh}.  %
Note that in the portion of the diagram to the right of the curve
$x=\xi(t)$ (\ie, to the right of the moving kink in the velocity
profile), spacetime is static.  For $t\to -\infty$, the geometry is
Minkowskian and the worldlines tend to approach straight lines with
slope $1/c$.%

The sound ray that generates the event horizon corresponds to a
finite\footnote{It is easy to check this explicitly using
(\ref{UH-fin}), given the asymptotic behaviours of $\bar{v}$ and
$\xi$.} value $U_{\rm H}$ of the coordinate $U$.  Hence, in this
situation an event horizon always exists.  This is also clear from
the fact that the vertical half-line $x=0$, $t>t_{\rm H}$ in
figure~\ref{F:tx-bh} is an apparent horizon.%

\subsection{Critical black hole}%
\label{subsec:cbh-event}%

\begin{figure}[htbp]%
\vbox{ \hfil \scalebox{0.50}{ {\includegraphics{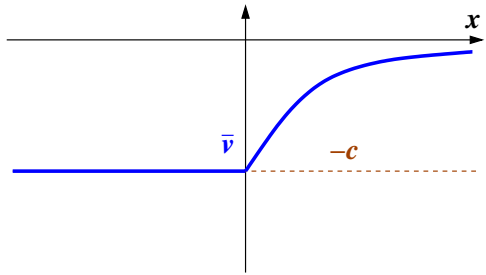}} }\hfil}%
\bigskip%
\caption{%
The static velocity profile $\bar{v}(x)$ for a critical black hole.%
} \label{F:v-cbh}%
\end{figure}

The function $\bar{v}(x)$ behaves as shown in figure~\ref{F:v-cbh}.
Regarding the right side of the profile, $x>0$, it is
indistinguishable from the profile of a non-extremal black hole
(figure~\ref{F:v-bh}).  %
%

\subsubsection{Finite time}%
\label{paragraph:cbh-ft}%

When the function $\xi(t)$ is of the form (\ref{appxi-bh}), that is,
when the apparent horizon is formed at a finite amount of laboratory
time, the situation is exactly the same as for the non-extremal black
hole discussed above.%

\subsubsection{Infinite time}%
\label{paragraph:cbh-it}%

\begin{figure}[htbp]%
\vbox{ \hfil \scalebox{0.50}{ {\includegraphics{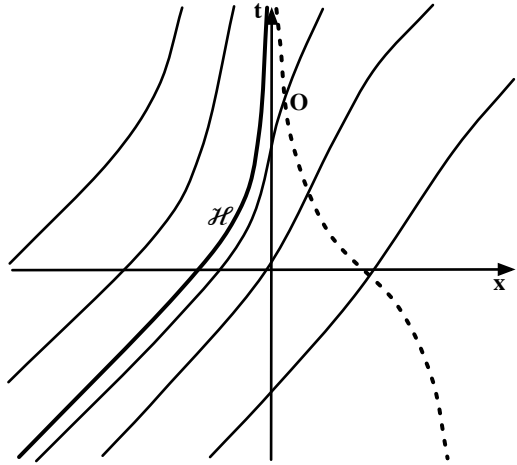}} }\hfil}%
\bigskip%
\caption{%
The worldlines of right-moving sound rays in the spacetime
describing the formation of a critical black hole. The
apparent horizon is just an asymptotic point at $x=0$, $t\to
+\infty$.  The event horizon, when it exists, corresponds to a
line like the thick solid one.  The worldline of the kink
is dashed. This situation is qualitatively identical to the case of the
formation of an extremal black hole in an infinite laboratory time.%
} \label{F:tx-cbh}%
\end{figure}

Consider now that the sonic point is approached in an infinite
amount of time, so the function $\xi(t)$ behaves as in
figure~\ref{F:xi-cbh}.  The worldlines of right-moving sound rays
are shown in figure~\ref{F:tx-cbh}.  %
As in the formation of the non-critical black hole, the portion of
the diagram to the right of the curve $x=\xi(t)$ (\ie, to the right
of the moving kink in the velocity profile) corresponds to a static
spacetime, and for $t\to -\infty$ the geometry is Minkowskian ---
the worldlines tend to approach straight lines with slope $1/c$.
However, now the apparent horizon is just the asymptotic point
located at $x=0$, $t\to +\infty$, and in order to establish whether
an event horizon does, or does not, exist one must perform an actual
calculation of $U_{\rm H}$ for the ``last'' ray that crosses the
kink.  The expression for $U_{\rm H}$ is again obtained from
equation (\ref{UU}), noticing that now  $t_0=+\infty$
along the generator of the would-be horizon, so%
\begin{equation}%
U_{\rm H}=\lim_{t_0\to +\infty}\left(t_0
+\frac{1}{c}\int_{-\infty}^{t_0}\d
t\,\bar{v}\left(\xi(t)\right)\right)\;.%
\label{barU-cbh}%
\end{equation}%
The necessary and sufficient condition for the event horizon to
exist is that the limit on right hand side of equation
(\ref{barU-cbh}) be finite.  The integrand on right hand side of
(\ref{barU-cbh}) can be approximated, for $t\to t_0\to +\infty$, as
$-c+\kappa\,\xi(t)$, while for $t\to -\infty$ it just approaches
zero.  Hence $U_{\rm H}$ is, up to a finite constant, equal to
$\kappa/c$ times the integral of $\xi$, evaluated at $t \to
+\infty$.  Here we must distinguish between the exponential
behaviour and the power law --- cases (i) and (ii).  In the former
$U_{\rm H}$ is finite, trivially.  For the power law, it turns out
that $U_{\rm H}$ is finite iff $\nu>1$.%

\subsection{Extremal black hole}%
\label{subsec:ebh-event}%

\begin{figure}[htbp]%
\vbox{ \hfil \scalebox{0.50}{ {\includegraphics{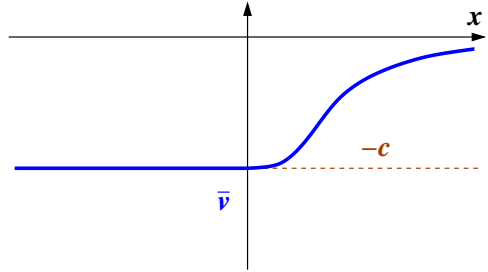}} }\hfil}%
\bigskip%
\caption{%
The static velocity profile $\bar{v}(x)$ for an acoustic extremal
black hole.%
} \label{F:v-ebh}%
\end{figure}

The typical spatial profile function $\bar{v}(x)$ for an extremal
black hole is plotted in figure~\ref{F:v-ebh}.  %
For $x$ approaching zero from positive values we can write%
\begin{equation}%
\bar{v}(x)=-c+\mu\,x^2+{\cal O}(x^3)\;,%
\label{appv-ebh}%
\end{equation}%
where $\mu>0$ is a constant.  As far as dynamics is concerned, we
must distinguish the cases in which the apparent horizon is formed
at finite laboratory time $t_{\rm H}$, and in an infinite time
(\ie, for $t\to +\infty$).%

\subsubsection{Finite time}%
\label{paragraph:ebh1}%

\begin{figure}[htbp]%
\vbox{ \hfil \scalebox{0.50}{{\includegraphics{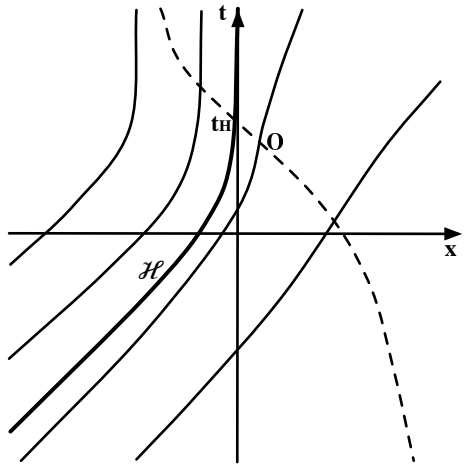}} }\hfil}%
\bigskip%
\caption{%
The worldlines of right-moving sound rays in the spacetime
describing the formation of an extremal black hole in a finite
laboratory time.  The thick solid and dashed lines represent,
respectively, the event horizon and the worldline of the kink.%
} \label{F:rays-ex-fin}%
\end{figure}

The function $\xi(t)$ is of the type shown in figure~\ref{F:xi-bh},
and the worldlines of right-going sound rays are sketched in
figure~\ref{F:rays-ex-fin}.  %
The event horizon always exists.%

\subsubsection{Infinite time}%
\label{paragraph:ebh2}%

The function $\xi(t)$ is as shown in figure~\ref{F:xi-cbh}, and the
worldlines of right-going signals are shown in
figure~\ref{F:tx-cbh}.
As in the case of the formation of a critical black hole, the apparent
horizon forms only asymptotically, for $x=0$ and $t\to +\infty$, so
the event horizon exists iff $U_{\rm H}$, given by equation
(\ref{barU-cbh}), has a finite value. Using the expansion
(\ref{appv-ebh}) in equation (\ref{barU-cbh}), one finds that $U_{\rm
H}$ is always finite when $\xi(t)$ is asymptotically exponential.  On
the other hand, for a power law, the event horizon exists iff
$\nu>1/2$.  (Note that the critical value of the exponent, $\nu =
1/2$, is now \emph{not} the same as for the critical black hole,
$\nu=1$.)%

\subsection{Double-sided black hole configurations}%
\label{subsec:2cbh-event}%

The configurations we have analyzed until now are the simplest
from a purely mathematical point of view.  However, having
in mind acoustic analogue geometries reproducible in a
one-dimensional pipe in the laboratory, it is more sensible to
consider double-sided configurations.  By this we mean that, after
passing (or approaching) the sonic/supersonic regime at $x=0$, and
traversing an interval of width $I\geq 0$, the fluid again goes
back to a subsonic regime as $x \to -\infty$.%

\begin{figure}[htbp]%
\vbox{ \hfil \scalebox{0.50}{ {\includegraphics{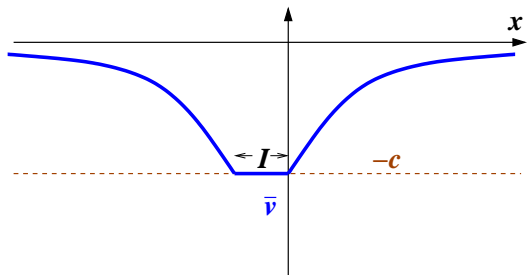}} }\hfil}%
\bigskip%
\caption{%
The static velocity profile $\bar{v}(x)$ for a double-sided critical
black hole. $I$ indicates the size of the internal flat segment.%
} \label{F:flat-spike}%
\end{figure}

\begin{figure}[htbp]%
\vbox{ \hfil \scalebox{0.50}{ {\includegraphics{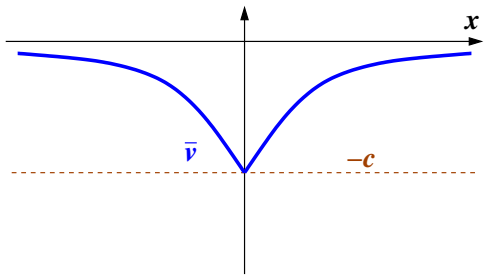}} }\hfil}%
\bigskip%
\caption{
The static velocity profile $\bar{v}(x)$ for a double-sided
critical black hole with zero ``thickness'' ($I=0$).%
} \label{F:spike}%
\end{figure}

\begin{figure}[htbp]%
\vbox{ \hfil \scalebox{0.50}{{\includegraphics{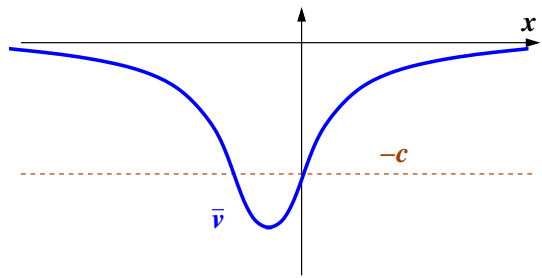}} }\hfil}%
\bigskip%
\caption{%
The static velocity profile $\bar{v}(x)$ for a double-sided
non-extremal black hole. In the companion paper \cite{companion}
we refer to this configuration as a black hole-white hole
configuration.%
} \label{F:v-bh-wh}%
\end{figure}

\begin{figure}[htbp]%
\vbox{ \hfil \scalebox{0.50}{{\includegraphics{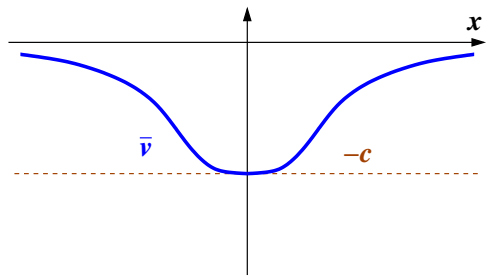}}} \hfil}%
\bigskip%
\caption{%
The static velocity profile $\bar{v}(x)$ for a double-sided
extremal black hole with zero thickness ($I=0$).%
} \label{F:v-ex-bh}%
\end{figure}

Consider for example functions $\bar{v}(x)$ such that
$\bar{v}(x)=-c$ for $-I \leq x \leq 0 $,%
\begin{equation}%
\lim_{x\to -I^{-}}\frac{\d \bar{v}(x)}{\d x}\neq 0\;,%
~~~~~
\lim_{x\to 0^{+}}\frac{\d \bar{v}(x)}{\d x}\neq 0\;,%
\end{equation}%
and which outside the interval $-I \leq x \leq 0 $ tend monotonically
to zero as $|x|$ increases (see figures~\ref{F:flat-spike} and
\ref{F:spike}).  %
The corresponding fluid configuration represents what could be
called a static ``double-sided critical black hole''.  The formation
of such a configuration can be modelled by the velocity
function%
\begin{equation}%
v(x,t)=\left\{\begin{array}{lll}%
\bar{v}(x)&\mbox{if}& x\geq\xi(t)\;,%
\\ %
\bar{v}(\xi(t))&\mbox{if}&  -I -\xi(t) \leq x \leq \xi(t)\;,%
\\ %
\bar{v}(x)&\mbox{if}& x \leq -I -\xi(t)\;,%
\end{array}\right.%
\label{velocity-dsc}%
\end{equation}%
with $\xi$ a monotonically decreasing function of $t$, and $\bar{v}$
as above.  Accordingly, the differential equation for right-going
sound rays also splits:%
\begin{equation}%
\frac{\d x}{\d t}=\left\{\begin{array}{lll}%
c+\bar{v}(x)&\mbox{if}& x \geq \xi(t)\;.%
\\ %
c+\bar{v}(\xi(t))&\mbox{if}&  -I -\xi(t) \leq x \leq \xi(t)\;,%
\\ %
c+\bar{v}(x)&\mbox{if}& x \leq -I -\xi(t)\;.%
\end{array}\right.%
\label{diffeq-dsc}%
\end{equation}%
Geometries associated with the formation of non-extremal and
extremal black holes can be constructed in the same way; see
figures~\ref{F:v-bh-wh} and \ref{F:v-ex-bh} for plots of
the respective $\bar{v}$ functions.  %
%

\subsubsection{Finite time}%
\label{paragraph:2cbh1}%

\begin{figure}[htbp]%
\vbox{ \hfil\scalebox{0.50}{{\includegraphics{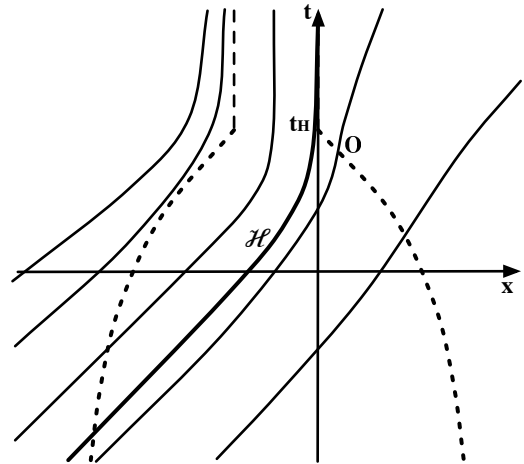}}} \hfil}%
\bigskip%
\caption{%
The worldlines of right-moving sound rays in the spacetime
describing the formation of a double-sided critical black hole in
a finite laboratory time. The event horizon is represented by
the thick solid line, while the worldlines of the kinks are
dashed.
} \label{F:rays-spike-fin}%
\end{figure}

The function $\xi(t)$ is of the type illustrated in
figure~\ref{F:xi-bh}.  The behaviour of right-going sound rays is
shown in figure~\ref{F:rays-spike-fin}.  %
The apparent horizon is the half-line $x=0$, $t>0$, and the event
horizon always exists.%

\subsubsection{Infinite time}%
\label{paragraph:2cbh2}%

\begin{figure}[htbp]%
\vbox{ \hfil\scalebox{0.50}{{\includegraphics{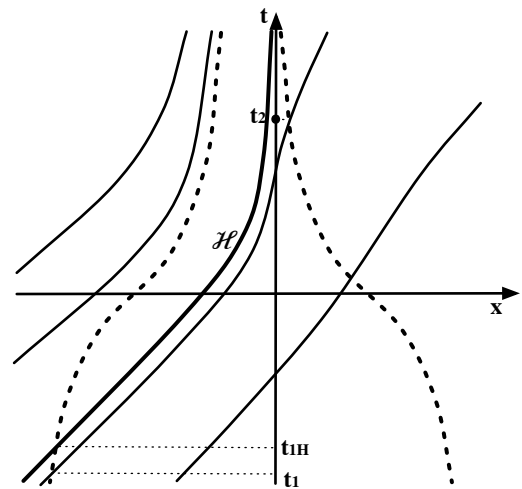}}} \hfil}%
\bigskip%
\caption{%
The worldlines of right-moving sound rays in the spacetime
describing the formation of a double-sided critical black hole in
an infinite laboratory time. The apparent horizon is the
asymptotic point $x=0$, $t\to +\infty$. The event horizon (when it
exists) is represented by the thick solid line, while the
worldlines of the kinks are dashed.
} \label{F:rays-spike-inf}%
\end{figure}

Let us now consider a function $\xi(t)$ of the type illustrated in
figure~\ref{F:xi-cbh}.  The behaviour of right-going sound rays is
shown in figure~\ref{F:rays-spike-inf}.  %
For these particular configurations, whether an event horizon does,
or does not, actually exist is now a rather tricky issue. The
asymptotic behaviour of the function $\xi(t)$ at $t \to -\infty$
ensures that all the right-going rays start to the left of the
right-moving kink (\ie, the one in the region $x<0$), then catch up
with it, and begin to propagate through the intermediate region at a
velocity $c+\bar{v}(\xi(t))$ that depends only on $t$.  Before they
reach the point $x=0$, such rays might be overtaken by the
right-moving kink, but only to start the chase again. After
several mutual overtakings (if the function $\xi(t)$ is sufficiently
complicated), the rays will always make an ultimate overtaking of
the right-moving kink, embarking upon a final encounter with the
left-moving kink on the right (\ie, in the region $x>0$). Let us
denote by $t_1$ the time of such a last crossing of the right-moving
kink, so the corresponding event is $(-\xi(t_1)-I,t_1)$. Also, let
us denote by $t_2$ the time at which the same ray crosses the kink
on the right, so that the corresponding event is $(\xi(t_2),t_2)$.
{From} equation (\ref{diffeq-dsc}) we directly obtain  the relation%
\begin{equation}%
\xi(t_1)+I=-\xi(t_2)+\int_{t_1}^{t_2}\d t\left[c
+\bar{v}\left(\xi(t)\right)\right]\;.%
\label{ziopollo}%
\end{equation}%
between $t_2$ and $t_1$.  (When the ray crosses the right-moving
kink more than once, equation (\ref{ziopollo}) will be satisfied by
more than one value of $t_1$ for any given $t_2$.  In order to
avoid cumbersome notation, we shall simply denote by $t_1$ the
largest of these roots, corresponding to the last crossing.) Then a
necessary and sufficient condition for the existence of an event
horizon is that, for $t_2\to +\infty$, $t_1$ tends to a finite value,
say $t_{1{\rm H}}$.  This guarantees that any right-going ray that
last crosses the left kink at a time greater than $t_{1{\rm H}}$
does not reach the region $x>0$ (as ray-crossing cannot occur
under the working hypothesis of this paper).%

Applying this condition straightforwardly in order to see whether
the event horizon exists is not easy.  Indeed, that would require us to
evaluate the integral in equation (\ref{ziopollo}) for a generic,
finite value of $t_2$, then solve for $t_1$ as a function of $t_2$.
It is easier to use one of the following two alternative
strategies:%
\begin{enumerate}%
\item Instead of asking whether the event horizon exists, one can ask
whether the event horizon {\em does not\/} exist.  A necessary and
sufficient condition for this is that, for $t_2\to +\infty$, also
$t_1\to +\infty$.  In such a case, we can insert the asymptotic
expansions (\ref{appv-bh}) or (\ref{appv-ebh}) into equation
(\ref{ziopollo}) to get%
\begin{equation}%
\xi(t_1)+I\sim -\xi(t_2)+\kappa\int_{t_1}^{t_2}\d t\,\xi(t)%
\label{appziopollo-ne}%
\end{equation}%
for a non-extremal black hole, and%
\begin{equation}%
\xi(t_1)+I\sim -\xi(t_2)+\mu\int_{t_1}^{t_2}\d t\,\xi(t)^2%
\label{appziopollo-e}%
\end{equation}%
for an extremal one.  Plugging into these expressions the different
asymptotic behaviours of the function $\xi(t)$, one can explicitly
solve for $t_1$ as a function of $t_2$ for large values of the
latter, and check whether $t_1$ does, or does not, tend to infinity
when $t_2\to +\infty$.%
\item Setting $t_2=+\infty$ into (\ref{ziopollo}), one obtains%
\begin{equation}%
\xi(t)+I=\int_{t}^{+\infty}\d t'\left[c
+\bar{v}\left(\xi(t')\right)\right]\;.%
\label{xiHinf}%
\end{equation}%
It is possible to show\footnote{We omit the somewhat delicate
proof of this statement in order not to overburden the
presentation.} that the event horizon exists if and only if equation
(\ref{xiHinf}) possesses an odd number of finite
solutions.\footnote{Note that, if this criterion is satisfied,
$t_{1{\rm H}}$ is the solution of equation (\ref{xiHinf}) with the
largest value.}  In order to establish whether this is the case, it
is convenient to define the function of $t$%
\begin{equation}%
f(t):=\int_t^{+\infty}\d t'\,\left[c+\bar{v}(\xi(t'))\right]\;,%
\label{integral}%
\end{equation}%
whose points of crossing with $\xi(t)+I$ correspond to the solutions
of equation (\ref{xiHinf}).  Of course, for $f$ to be well defined
(and therefore for solutions of (\ref{xiHinf}) to exist at all) one
needs the integral defining it to be convergent.  For asymptotically
($t \to +\infty$) exponential and power-law behaviours of $\xi(t)$
this happens in the cases already described.  Now, whenever $f$ is
well defined, it is clearly a monotonically decreasing function,
because the integrand in equation (\ref{integral}) is always
strictly positive.  For $t \to -\infty$, $f(t)$ is just equal to the
integral of the function $c+\bar{v}(\xi(t))$ evaluated at $t$, up to
a finite constant.  In this limit, $c+\bar{v}(\xi(t))\to c$ so we
can write, for $t\to -\infty$:%
\begin{equation}%
f(t)\sim -c\,t+\mbox{const}\;.%
\label{kt}%
\end{equation}%
Given the condition $\displaystyle\lim_{t \to -\infty} \dot \xi =
0$, it is clear that the function $f(t)$ is always greater than
$\xi(t)+I$ for $t\to -\infty$.  On the other hand, for $t\to
+\infty$, the asymptotic behaviour of $f(t)$ is obtained by
expanding $\bar{v}$ in (\ref{integral}), which gives%
\begin{equation}%
f(t)\sim \kappa\int_t^{+\infty}\d t'\,\xi(t')\;%
\label{f+inf}%
\end{equation}%
for critical (and non-extremal) black holes and%
\begin{equation}%
f(t)\sim \mu \int_t^{+\infty}\d t'\,\xi^2(t')\;%
\label{f+inf+extr}%
\end{equation}%
for extremal ones.  If, for $t\to +\infty$, $f(t)$ is smaller
(greater) than $\xi(t)+I$, then equation (\ref{xiHinf}) has an odd
(even) number of finite solution, and the event horizon does (does
not) exist.  Note that, if $f(t)\sim \xi(t)+I$, one must analyse
subdominant terms in the asymptotic behaviour of $\xi(t)$ in order
to draw any conclusion.%
\end{enumerate}%
With either method, we find that for $I \neq 0$ the existence of an
event horizon in double-sided configurations follows the same rules
as in the previously analysed one-sided configurations.  When
$I=0$ however, it is more difficult to have an event horizon in a
double sided configuration, and in general, one has to increase the
{\em rapidity\/} with which one approaches the sonic regime.  More
specifically, for a critical black hole and an asymptotically
exponential $\xi(t)$, the event horizon exists if $\kappa_{\rm
D}>\kappa$, but not when $\kappa_{\rm D}<\kappa$, while for a power
law there is no horizon.  For an extremal black hole and an
exponential $\xi(t)$ the horizon always exists, but in the case of a
power law it does not exist if $\nu<1$, and it exists for $\nu\geq
1$, with the additional condition $B<1/\mu$ for the particular
value $\nu=1$.  For a critical black hole with asymptotically
exponential $\xi(t)$ and $\kappa_{\rm D}=\kappa$, as well as for an
extremal black hole with a power law and $\nu=1$, $B=1/\mu$, the
asymptotic analysis is not sufficient and one must take into account
also subdominant terms in the expansion of $\xi(t)$ for $t\to
+\infty$.%

\section{Asymptotic redshift relations}%
\label{sec:creation}%
\setcounter{equation}{0}%

For those situations in which an event horizon exists, we now
find the asymptotic relation between $u$ and $U$ for rays close to
the horizon generator.  We also briefly discuss the implications of
such a relation for quasi-particle creation in the various cases of
interest.%

\subsection{Non-extremal black hole}%
\label{subsec:bh}%

Consider a sound ray corresponding to a value $U<U_{\rm H}$. For $U$
very close to $U_{\rm H}$, $t_0$ is very close to $t_{\rm H}$, and we can
use the approximation (\ref{appxi-bh}) for $\xi(t)$. Furthermore, we
can approximate $\bar{v}$ as in (\ref{appv-bh}), so equation
(\ref{U-fin}) gives%
\begin{equation}%
U=U_{\rm H}+\frac{\lambda}{c}\,(t_0-t_{\rm H})+{\cal O}([t_0-t_{\rm H}]^2)\;.%
\label{Ut0-bh}%
\end{equation}%
This provides us with the link between $U$ and $t_0$.%

In order to link $t_0$ with $u$, consider the integral on the right
hand side of equation (\ref{uu}).  For $x\to +\infty$, the integrand
function vanishes, while near $\xi(t_0)$ it can be approximated by
$-c/(\kappa x)$.  Then the integral is just given by the difference of
the corresponding integrals evaluated at $x=+\infty$ and
$x=\xi(t_0)$, respectively, up to a possible finite positive
constant.  This gives\footnote{This result could also have been
obtained by noticing that the corresponding part of the worldline lies
into a static portion of spacetime, for which one can simply use the
representative profile for $\bar{v}$ given in
reference~\cite{companion}.  Using equation (4.2) from that paper we
have%
\begin{equation}%
u=t_0-t_{\rm H}-\frac{\xi(t_0)}{c}-\frac{1}{\kappa}\,\ln\left|1
-\e^{-2\kappa\xi(t_0)/c}\right|\;.%
\label{t0u1-bh}%
\end{equation}%
Expanding, we find again equation (\ref{porcoqua}), to the leading
order in $t_0-t_{\rm H}$.}%
\begin{equation}%
-\lambda\,(t_0-t_{\rm H})\sim \mbox{const}\,\e^{-\kappa u}\;.%
\label{porcoqua}%
\end{equation}%
Together with equation (\ref{Ut0-bh}), this leads to%
\begin{equation}%
U\sim U_{\rm H}-\mbox{const}\,\e^{-\kappa u}\;.%
\label{Uu'-bh}%
\end{equation}%
This relation between $U$ and $u$ is exactly the one found by
Hawking in his famous analysis of particle creation by a collapsing
star~\cite{hawking}. It is by now a standard result that this
relation implies the stationary creation of particles with a Planckian
spectrum at temperature $\kappa/(2\pi)$~\cite{bd,hu}.%

\subsection{Critical black hole}%
\label{subsec:cbh}%

For a critical black hole, the results are very different
according to whether the sonic regime is attained in a finite or
an infinite laboratory time.%

\subsubsection{Finite time}%
\label{subsubsec:cbh-ft}%

The calculation of the relation between $U$ and $u$ is exactly equal
to the one presented for the non-extremal black hole case.  The two
geometries coincide everywhere to the right of the apparent horizon
and cannot be distinguished by the quasi-particle production
observed at $x \to +\infty$.%

\subsubsection{Infinite time}%
\label{subsubsec:cbh-it}%

Let us suppose that we are in a situation in which the event horizon
exists, so $U_{\rm H}$ is finite.  For another right-moving sound
ray that corresponds to a value $U<U_{\rm H}$ we find, combining
equations (\ref{UU}) and (\ref{barU-cbh}),%
\begin{equation}%
U=U_{\rm H}-\frac{\xi(t_0)}{c}-\frac{1}{c}\int_{t_0}^{+\infty}\d
t\left[c+\bar{v}\left(\xi(t)\right)\right]\;.%
\label{deltaU-cbh}%
\end{equation}%
In the integration interval, $\xi(t)$ is close to zero, so
equation (\ref{deltaU-cbh}) can be approximated as%
\begin{equation}%
U\sim U_{\rm
H}-\frac{\xi(t_0)}{c}-\frac{\kappa}{c}\int_{t_0}^{+\infty}\d
t\,\xi(t)\;,%
\label{deltaU'-cbh}%
\end{equation}%
where the expansion (\ref{appv-bh}) has been used.  Equation
(\ref{deltaU'-cbh}) gives%
\begin{equation}%
U\sim
U_{\rm H}-\frac{A}{c}\left(1+\frac{\kappa}{\kappa_{\rm D}}\right)
\e^{-\kappa_{\rm D}\,t_0}%
\label{deltaUexp}%
\end{equation}%
for an asymptotically exponential $\xi$, and%
\begin{equation}%
U\sim U_{\rm H}-\frac{\kappa\,B}{(\nu-1)\,c}\,t_0^{-(\nu-1)}%
\label{deltaUnu}%
\end{equation}%
for a power law with $\nu>1$.%

For the link between $t_0$ and $u$ we obtain%
\begin{equation}%
u\sim t_0-\frac{1}{\kappa}\,\ln\xi(t_0)\;,%
\label{tu}%
\end{equation}%
as one can easily check inserting the appropriate asymptotic expansions
into equation (\ref{uu}).\footnote{One could again also use
equation (4.2) from reference \cite{companion}; this leads to
equation (\ref{t0u1-bh}) which, expanded, gives equation (\ref{tu}).
The result holds, however, independently of the details of
$\bar{v}(x)$.}  Using equation (\ref{tu}) into equations
(\ref{deltaUexp}) and (\ref{deltaUnu}) we find%
\begin{equation}%
U\sim U_{\rm
H}-\mbox{const}\,\exp\left(-\frac{\kappa\,\kappa_{\rm D}}{\kappa
+\kappa_{\rm D}}\,u\right)%
\label{Uuexp}%
\end{equation}%
for the exponential case, and%
\begin{equation}%
U\sim U_{\rm H}-\mbox{const}\,u^{-(\nu-1)}%
\label{Uunu}%
\end{equation}%
for a power law with $\nu>1$. (Remember that for $\nu\leq1$ the event
horizon does not form.)%

It is interesting to compare equations (\ref{Uuexp}) and (\ref{Uunu})
with the corresponding one for the non-critical black hole, equation
(\ref{Uu'-bh}).  Whereas the latter is basically independent of the
details of the black hole formation (which only appear in the
multiplicative constant), the relation between $U$ and $u$ in the
critical case is not universal, but depends on the dynamical evolution.
Even for an asymptotically exponential $\xi(t)$, which leads to an
exponential dependence on $u$, the coefficient in the exponent is not
universal as in equation (\ref{Uu'-bh}), but depends on dynamics
through the parameter $\kappa_{\rm D}$.  This is not difficult to
understand looking back at the way in which equations (\ref{Uu'-bh})
and (\ref{Uuexp}) have been derived.  For equation (\ref{Uu'-bh}), the
exponential dependence was introduced relating $t_0$ with $u$, which
only involves sound propagation in the final static region and cannot,
therefore, be affected by dynamics.  On the other hand, when deriving
equation (\ref{Uuexp}) it is sound propagation in the initial,
dynamical, regime that introduces the exponential (in the particular
case of an asymptotically exponential $\xi$); hence, it is not
surprising that the final result keeps track of the dynamical
evolution. However, it is interesting to note that in the limit
$\kappa_{\rm D}\to +\infty$ equations (\ref{Uu'-bh}) and (\ref{Uuexp})
coincide. This limit corresponds to a very rapid approach towards the
formation of an otherwise-never-formed (in finite time) apparent
horizon. Regarding the creation of quasi-particles, this situation is
operationally indistinguishable from the actual formation of the sonic
point.  However, this ``degeneracy'' might be accidental,
given that the origin of the exponential relation is very
different in the two cases.%

\subsection{Extremal black hole}%
\label{subsec:ebh}%

As for the case of a critical black hole, we must distinguish
between a finite and an infinite time of formation of the event
horizon.%

\subsubsection{Finite time}%
\label{subsubsec:ebh1}%

For a sound ray close to the one that generates the horizon,
equation (\ref{U-fin}) still holds. However, now one must use the
expansion (\ref{appv-ebh}) when approximating the integrand thus
obtaining%
\begin{equation}%
U\sim U_{\rm H}+\frac{\lambda}{c}\,(t_0-t_{\rm H})
+{\cal O}([t_0-t_{\rm H}]^3)\;.%
\label{deltaUt0-ebh1}%
\end{equation}%
Using again the approximation (\ref{appv-ebh}) in the evaluation of
the integral on the right hand side of equation (\ref{uu}) one
finds%
\begin{equation}%
t_0\sim t_{\rm H}-\lambda\,\mu\,u^{-1}\;.%
\label{ut-ebh1}%
\end{equation}%
Finally,%
\begin{equation}%
U\sim U_{\rm H}-\frac{1}{\mu\,c}\,u^{-1}\;.%
\label{Uu-ebh1}%
\end{equation}%
Interestingly, this is the same relation that one finds for the
gravitational case \cite{lrs}. In particular, this implies that
finite time collapse to form an extremal black hole will \emph{not}
result in a Planckian spectrum of quasi-particles~\cite{lrs}.
This is completely compatible with the standard GR analysis, and is
one of the reasons why extremal and non-extremal black holes are
commonly interpreted as belonging to completely different
thermodynamic sectors~\cite{LRS2}.%

\subsubsection{Infinite time}%
\label{subsubsec:ebh2}%

Assuming that the event horizon exists, we can again apply equation
(\ref{deltaU-cbh}) and use the approximation (\ref{appv-ebh}) in
order to find the relation between $U$ and $t_0$.  The results
are, for an asymptotically exponential $\xi(t)$:%
\begin{equation}%
U\sim U_{\rm H}-\frac{A}{c}\,\e^{-\kappa_{\rm D}\,t_0}\;;%
\label{Utexp-ebh}%
\end{equation}%
for a power law with $1/2<\nu<1$:%
\begin{equation}%
U\sim U_{\rm H}-\frac{B}{c}\,t_0^{-\nu}\;;%
\label{Utnu<1}%
\end{equation}%
for a power law with $\nu=1$:%
\begin{equation}%
U\sim U_{\rm H}-\frac{B}{c}\left(1+\mu\,B\right)t_0^{-1}\;;%
\label{Utnu=1}%
\end{equation}%
for a power law with $\nu>1$:%
\begin{equation}%
U\sim U_{\rm H}-\frac{\mu\,B^2}{\left(2\,\nu
-1\right)c}\,t_0^{-2\,\nu+1}\;.%
\label{Utnu>1}%
\end{equation}%
Using the appropriate expansions in equation
(\ref{t-t}),\footnote{Or equation (4.14) in
reference~\cite{companion}.} one obtains the relation
between $t_0$ and $u$:%
\begin{equation}%
u\sim t_0+\frac{1}{\mu\,\xi(t_0)}\;.%
\label{tu-ebh}%
\end{equation}%
For an asymptotically exponential $\xi(t)$ this becomes%
\begin{equation}%
u\sim \frac{1}{\mu\,A}\,\e^{\kappa_{\rm D}\,t_0}\;.%
\label{tuexp-ebh}%
\end{equation}%
For a power law, one must again distinguish between three cases;
for $1/2<\nu<1$:%
\begin{equation}%
u\sim t_0\;;%
\label{utnu<1}%
\end{equation}%
for $\nu=1$:%
\begin{equation}%
u\sim \left(1+\frac{1}{\mu\,B}\right)t_0\;;%
\label{utnu=1}%
\end{equation}%
for $\nu>1$:%
\begin{equation}%
u\sim \frac{1}{\mu\,B}\,t_0^\nu\;.%
\label{utnu>1}%
\end{equation}%
Putting together equations (\ref{Utexp-ebh}) and (\ref{tuexp-ebh})
one finds the relationship between $U$ and $u$ for the exponential
case:%
\begin{equation}%
U\sim U_{\rm H}-\frac{1}{\mu\,c}\,u^{-1}\;.%
\label{Uuexp-ebh}%
\end{equation}%
For the power law one finds from equations
(\ref{Utnu<1})--(\ref{Utnu>1}) and (\ref{utnu<1})--(\ref{utnu>1}),
for $1/2<\nu<1$:%
\begin{equation}%
U\sim U_{\rm H}-\frac{B}{c}\,u^{-\nu}\;;%
\label{Uunu<1}%
\end{equation}%
for $\nu=1$:%
\begin{equation}%
U\sim U_{\rm H}-{\left(1+\mu\,B\right)^2}{\mu\,c}\,u^{-1}\;;%
\label{Uunu=1}%
\end{equation}%
and finally, for $\nu>1$:%
\begin{equation}%
U\sim U_{\rm H}-\frac{\left(\mu\,B\right)^{1/\nu}}{\left(2\,\nu
-1\right)\mu\,c}\,u^{-(2-1/\nu)}\;.%
\label{Uunu>1}%
\end{equation}%
In all these cases, quasi-particle production is neither
universal, nor Planckian.%

\subsection{Double-sided black hole configurations}%
\label{subsec:2cbh}%

It is not difficult to prove that in the formation, in a
finite amount of time, of double-sided non-extremal black holes,
double-sided extremal black holes, and double-sided critical black
holes, the asymptotic relation between $U$ and $u$ is identical to
that calculated in the corresponding subsections above.  The
amount and features of quasi-particle creation are then the
same. We will demonstrate this in detail for the case of a
double-sided critical black hole, and then proceed to consider the
situation in which the formation takes place in an infinite amount
of time.%

\subsubsection{Finite time}%
\label{subsubsec:2cbh1}%

Using the same notation as in section~\ref{paragraph:2cbh2},
let us call $t_1$ the largest of the $t_1$'s that satisfy
equation (\ref{ziopollo}), so $t_1$ is the time at which a
right-going ray last crosses the kink on the left.  There will be
some regular relationship between $U$ and $t_1$, expressed by
some differentiable function $f$, so that we can write
$U=f(t_1)$. For the event horizon to exist, the corresponding
$U$ must be finite (equal to some value $U_{\rm H}$, say), so also
$t_{1{\rm H}}=f^{-1}(U_{\rm H})$ must be finite (as already
done in section~\ref{paragraph:2cbh2}, we denote by a suffix ``H''
the quantities that correspond to the horizon generator).%

For a ray very close to the horizon generator we have%
\begin{equation}%
U=U_{\rm H}+f(t_1)-f(t_{1{\rm H}})\sim U_{\rm
H}-\dot{f}(t_{1{\rm H}})\left(t_{1{\rm H}}-t_1\right)\;,%
\label{Uf}%
\end{equation}%
where a dot denotes the derivative with respect to $t$.  On the
horizon, $t_2=t_{\rm H}$ so equation (\ref{ziopollo}) reduces to%
\begin{equation}%
\xi(t_{1{\rm H}})+I=\int_{t_{1{\rm H}}}^{t_{\rm H}}\d t
\left[c
+\bar{v}\left(\xi(t)\right)\right]\;.%
\label{ziogallo}%
\end{equation}%
Subtracting (\ref{ziogallo}) from (\ref{ziopollo}) we obtain%
\begin{equation}%
\xi(t_1)-\xi(t_{1{\rm H}})=-\xi(t_2)
+\int_{t_1}^{t_{1{\rm H}}}\d t \left[c
+\bar{v}\left(\xi(t)\right)\right] -\int_{t_2}^{t_{\rm H}}\d
t\left[c
+\bar{v}\left(\xi(t)\right)\right]\;.%
\label{ziopollogallo}%
\end{equation}%
For a ray close to the horizon generator, $t_2$ is close to $t_{\rm
H}$, and $t_1$ close to $t_{1{\rm H}}$, so equation
(\ref{ziopollogallo}) gives, keeping only terms to the leading
order:%
\begin{equation}%
t_{1{\rm H}}-t_1\sim\frac{\lambda}{c
+\bar{v}\left(\xi(t_{1{\rm H}})\right)
+\dot{\xi}(t_{1{\rm H}})}\,(t_{\rm H}-t_2)\;.%
\label{eccola}%
\end{equation}%
Together, equations (\ref{Uf}) and (\ref{eccola}) provide a linear
link between $U$ and $t_2$.  Since the relationship between
$t_2$ and $u$ is exactly the same as the one between $t_0$ and
$u$ in equation (\ref{porcoqua}), the final result is again the one
expressed by (\ref{Uu'-bh}):
\[
U \sim U_{\rm H}-\mbox{const}\,\e^{-\kappa u}\;.
\]

\subsubsection{Infinite time}%
\label{subsubsec:2cbh2}%

Assuming that we are in a situation for which the event
horizon does indeed exist, we can subtract equation (\ref{xiHinf})
with $t_1\to t_{1{\rm H}}$ from equation (\ref{ziopollo}),
finding:%
\begin{equation}%
\xi(t_1)-\xi(t_{1{\rm H}})=-\xi(t_2)
+\int_{t_1}^{t_{1{\rm H}}}\d t\left[c
+\bar{v}\left(\xi(t)\right)\right]-\int_{t_2}^{+\infty}\d
t\left[c +\bar{v}\left(\xi(t)\right)\right]\;.%
\label{xi-xi}%
\end{equation}%
For a ray close to the horizon generator, $t_1$ is close to
$t_{1{\rm H}}$ and $t_2$ is large, so%
\begin{equation}%
t_{1{\rm H}}-t_1\sim \frac{\xi(t_1)+\kappa
\displaystyle\int_{t_2}^{+\infty}\d t\,\xi(t)}{c
+\bar{v}\left(\xi(t_{1{\rm H}})\right)
+\dot{\xi}(t_{1{\rm H}})}\;.%
\label{eccolaqui}%
\end{equation}%
For an asymptotically exponential $\xi(t)$ we find,
performing the integral,%
\begin{equation}%
t_{1{\rm H}}-t_1\sim \frac{A}{c
+\bar{v}\left(\xi(t_{1{\rm H}})\right)
+\dot{\xi}(t_{1{\rm H}})}\left(1
+\frac{\kappa}{\kappa_{\rm D}}\right)\e^{-\kappa_{\rm D} t_2}\;.%
\label{eccolali}%
\end{equation}%
Similarly, for a power law with $\nu>1$:%
\begin{equation}%
t_{1{\rm H}}-t_1\sim \frac{B\,\kappa}{(\nu-1)\left(c
+\bar{v}\left(\xi(t_{1{\rm H}})\right)
+\dot{\xi}(t_{1{\rm H}})\right)}\,t_2^{-(\nu-1)}\;.%
\label{eccolala}%
\end{equation}%
In both cases, the same results as in section~\ref{subsec:cbh},
equations (\ref{Uuexp}) and (\ref{Uunu}), follow.%

In short, the amount and characteristics of the quasi-particle
production calculated with the double-sided configurations are
exactly the same as those calculated with the simpler profiles in
the previous subsections except in two specific situations: The
double-sided critical black hole with $I=0$ (see
figure~\ref{F:spike}) and the double-sided extremal black hole. In
the critical case, only the asymptotically exponential behaviour
with $\kappa_{\rm D} > \kappa$ produces an event horizon and,
therefore, only then we can talk about a stationary and Planckian
creation of quasi-particles. In the extremal case the results
described in section~\ref{subsubsec:ebh2} only apply for $\nu
\geq 1$ (with the further condition $B < 1/\mu$ in the particular
case $\nu=1$), because otherwise the event horizon itself does not
exist.%

\section{Conclusions and discussion}%
\label{sec:discussion}%
\setcounter{equation}{0}%

In the present paper we have analyzed different dynamical black
hole-like analogue geometries with regard to their properties in terms of
quantum quasi-particle production.  We have taken several $(1+1)$-dimensional spacetimes (considered as externally fixed
backgrounds), and for each of them we (i) have calculated whether
it possesses an event horizon or not, and if the answer is
``yes'', (ii) have calculated the asymptotic redshift function that
characterizes the amount and properties of the late-time quasi-particle production.  In Table~\ref{table} the reader can
find a summary of all our results.%
%
\noindent%
\begin{table}%
\begin{center}
\small
\begin{tabular}{|c|c|c||c||c|c|}%
\hline\hline
\multicolumn{3}{|c||}{\em Black hole type\/} & {\em Horizon?\/} &
{\em Redshift\/} & {\em Equation}\\%
\hline\hline
{non-extremal} &\multicolumn{2}{c||}{finite time}&  {always} &
{exponential} & (\ref{Uu'-bh}) \\%
\hline
\hline%
{critical and } &\multicolumn{2}{c||}{finite time} & {always} &
{exponential} & (\ref{Uu'-bh}) \\%
\cline{2-6} {double-sided critical} & {infinite} & {exponential} &
{always} & {exponential} &  (\ref{Uuexp})\\%
\cline{3-6} {with $I\neq 0$} & {time}& {power law} & {for $\nu>1$} &
{power law} & (\ref{Uunu})\\%
\hline\hline%
{extremal and} &\multicolumn{2}{c||} {finite time} & {always} &
{power law} &
(\ref{Uu-ebh1}) \\
\cline{2-6}{double-sided extremal} & {infinite}& {exponential} &
{always} &
{power law} &  (\ref{Uuexp-ebh})\\%
\cline{3-6}{with $I\neq 0$} & {time}&{power law} & { for $\nu>1/2$}
&
{power law} & (\ref{Uunu<1})--(\ref{Uunu>1})\\%
\hline\hline%
{double-sided} & \multicolumn{2}{c||} {finite time} &
{always} & {exponential} &
(\ref{Uu'-bh}) \\%
\cline{2-6}{critical} & {infinite}&{exponential} & { for
$\kappa_{\rm D}>\kappa$} & {exponential} & (\ref{Uuexp})\\%
\cline{3-6}{with $I=0$} & {time}&{power law} & {never} &
{} & {}\\%
\hline\hline%
{double-sided} & \multicolumn{2}{c||} {finite time} &
{always} & {power law} & (\ref{Uu-ebh1}) \\%
\cline{2-6}{extremal} & {infinite} & {exponential} & {always} &
{power law} & (\ref{Uuexp-ebh})\\%
\cline{3-6}{with $I=0$} & {time} &{power law} & {for $\nu\geq 1$} &
{power law} & (\ref{Uunu=1})--(\ref{Uunu>1}) \\%
\hline\hline%
\end{tabular}%
\end{center}
\caption{This is a summary of the results found for the different 
configurations analyzed in the paper. To the table we have to add the
following comments: For double-sided extremal black hole with $I=0$,
with an infinite time of formation and an asymptotic power law for
$\xi(t)$, the horizon forms in the case $\nu=1$ if the further
condition $B<1/\mu$ holds.  For the double-sided configurations with
$I=0$ and an infinite time of formation, the asymptotic analysis is
not sufficient for drawing conclusions when $\xi(t)$ is asymptotically
exponential and $\kappa_{\rm D}=\kappa$, and when $\xi(t)$ is
asymptotically a power law and $\nu=1$, $B=1/\mu$.  For those cases,
one need also consider subdominant terms in $\xi(t)$, so the results
will depend on the details of formation.}%
\label{table}%
\end{table}%

The above results are pertinent to a purely mathematical 
model. Their physical relevance has to be assessed with respect to their
application to  both experimental reproduction of the analogue
Hawking radiation, and to the lessons they can provide concerning the 
possible behavior of black hole formation and evaporation in
semiclassical gravity. We now turn to separately consider these
two issues.%

\subsection{Experimental realizability}%
\label{sec:exp-real}%

The study carried on in this paper has identified several velocity
profiles that are potentially interesting for experiments. In particular
the critical black hole models seem worth taking into consideration 
in connection with the realizability of a
Hawking-like flux in the laboratory.  The creation of supersonic
configurations in a laboratory is usually associated with the
development of instabilities.  There are many examples of the latter 
in the literature; {\eg} in reference~\cite{volovik-ripplons} it was
shown that in an analogue model based on ripplons on the interface
between two different sliding superfluids (for instance, 
$^3$He-phase A and $^3$He-phase B), the formation of an ergoregion
would make the ripplons acquire an amplification factor that
eventually would destroy the configuration.  Therefore, this
analogue system, although very interesting in its own right, will prove to 
be useless in terms of detecting a Hawking-like flux.  However, by
creating, instead of an ergoregion, a critical configuration one
should be able, at least, to have a better control of the incipient instability, while at the same time producing a dynamically controllable
Hawking-like flux.%

Nevertheless, the actual realization of a critical configuration could
also appear as a problematic task for entirely different reasons. The corresponding velocity profiles are characterized by
discontinuities in the derivatives, so one might wonder whether they would be amenable to experimental construction, given that the continuum model
is only an approximation. Let us therefore discuss in some detail the
validity of the latter for realistic systems.%

The main difference between an ideal perfect fluid model and a realistic 
condensed matter analogue is due to the microscopic structure of the system considered. In particular, it is generic to have a length scale
$\delta$ which characterizes the breakdown of the continuum model 
($\delta$ is of the order of the intermolecular distance for an ordinary
fluid; of the coherence length for a superfluid; and of the healing length
for a Bose--Einstein condensate). In general, the viability of the analogue 
model requires one to consider distances $\Delta$ of order of at least a
few $\delta$, depending on the accuracy of the experiment performed. In
particular, wave propagation is well defined only for wavelengths larger 
than $\delta$ (generally with an intermediate regime, for wavelengths
between $\delta$ and $\Delta$, where the phenomena exhibit deviations
with respect to the predictions based on the continuum model).%

In general, a mathematical description based on the continuum model
contains details involving scales smaller than $\Delta$ (for
example, in the velocity profile). These details should, however, be
regarded as unphysical: They are present in the model, but do not 
correspond to properties  of the real physical system. In particular,
they cannot be detected experimentally, because this would require
{\eg} using wavelengths smaller than $\Delta$,  which do not behave
according to the predictions of the model (and for wavelengths smaller 
than $\delta$ do not even make physical sense).%

For the mathematical models considered in the present paper, all this
implies that one will not be able to distinguish, on empirical grounds,
between those cases for which the velocity profiles differ from each 
other only by small-scale details. In particular, double-sided
configurations with $I=0$ should be equivalent to configurations
with a small, but non-zero, thickness $I<\Delta$. Also, one
would  not be able to distinguish between two velocity profiles that 
differ only in a neighborhood $\Delta$ of $x=0$, one of which corresponds
to a critical black hole, while the other describes an extremal one.
In particular Hawking radiation will not distinguish between the models 
within each of these pairs.%

This fact would not be troublesome, had our analysis led to
identical results for the acoustic black holes of each pair. However,
this is not the case (see Table~\ref{table}).  But then what shall we 
see if we realize these models in a laboratory?%

In realistic situations, what is relevant for Hawking radiation is a
coarse-grained profile obtained by averaging over a scale of order
$\Delta$, thus neglecting the unphysical small scale details in $\bar{v}(x)$. 
This implies that as far as double-sided critical black holes are
concerned, the reliable results are those pertinent to the non-zero
thickness case ($I\neq0$). Similarly, since these extremal black holes are never exactly realizable in a laboratory (as this would require tuning the 
velocity profile on arbitrary small scales), only the predictions
based on the critical black hole mathematical model will survive
in an experimental setting. Indeed, the relevant surface gravity will
be defined by averaging the slope of the velocity profile over scales 
which are of order of $\Delta$.\footnote{The average is the one from
the right, since we know that it is the slope in the proximity of the
second kink that is responsible for the Hawking-like effect.} This  averaged surface gravity will be non-zero for both the critical and the
extremal black hole, but will be approximately equal to the surface
gravity at the horizon of the critical black hole, while it will obviously 
not coincide with the one of the extremal (which is zero).

\subsection{Hints for semiclassical gravity}\label{sec:sem-grav}
In the body of the paper we have used a terminology particularly
suitable to dealing with analogue models based on acoustics. Let us
now discuss the most relevant features of our findings
using a language more natural to GR.%

When the geometry associated with the formation of a spherically
symmetric black hole through classical gravitational collapse (as,
for example, in the Oppenheimer-Snyder
model~\cite{oppenheimer-snyder}) is described in terms of
Painlev\'e-Gullstrand~\cite{pg} coordinates (whose counterpart, in
the context of acoustic geometries, are the natural laboratory
coordinates $x$ and $t$), the apparent horizon forms in a
finite amount of coordinate time.  In this regard, the
Painlev\'e-Gullstrand time behaves similarly to the proper time
measured by a freely-falling observer attached to the surface
of the collapsing star. The non-extremal, non-critical
$(1+1)$-dimensional model analysed in this paper, captures the main
features of the formation of a (non-extremal) black hole. The
dynamical collapse is represented by the function $\xi(t)$ in our
calculations.  In the language of GR, we can think of $\xi(t)$ as
the radial distance between the surface of a collapsing star and its
Schwarzschild radius; $\xi(t_{\rm H})=0$ corresponds to the moment
in which the surface of the star enters its Schwarzschild radius,
and this moment corresponds to a finite time (which we took to be
$t_{\rm H}$).%

For this model we recovered Hawking's result that the formation of
(non-extremal) black holes causes the quantum emission towards
infinity of a stationary stream of radiation with a Planckian
spectrum, at temperature $\kappa/(2\pi)$.  The mechanism for
particle creation is somewhat ``more than dynamical'' as the
characteristics of the stationary stream of particles are
``universal'' and only depend on the properties of the geometry at
the horizon, $\kappa$, and not on any detail of the dynamical
collapse.  Indeed, for $\xi(t)$ given by equation
(\ref{appxi-bh}) --- apparent horizon formation in a finite amount
of time --- we have seen that asymptotic quasi-particle
creation does not depend even on the coefficient $\lambda$.  That
is, particle production does not depend on the velocity with which
the surface of the collapsing star enters its Schwarzschild radius.%

This picture leans toward the (quite standard) view that Hawking's
process is not just dynamical, but relies on the actual existence of
an apparent horizon and an ``ergoregion'' beyond it, able to absorb
the negative energy pairs~\cite{bd,parker}.  However, by analyzing
alternative models, in this paper we have seen two unexpected
things:%
\begin{description}%
\item[{\rm i)}] One can also produce a truly Hawking flux with
a temperature $\kappa/(2\pi)$ through the formation in a finite
amount of time 
 of either a single-sided critical black hole, or a double-sided critical black hole of finite ``thickness", or even one of zero ``thickness" (see figure~\ref{F:spike}).  This is an intriguing result, as in none of these cases there is an ``ergoregion'' beyond the apparent horizon, and in the last case there is just a single sonic point. (In the language of GR, this last configuration corresponds to stopping the collapse of a star at the very moment in
which its surface reaches the Schwarzschild
radius.)%
\item[{\rm ii)}] Moreover, one can also produce a stationary
and Planckian emission of quasi-particles by, instead of
actually forming the apparent horizon, just approaching its
formation asymptotically in time with sufficient rapidity ($\xi(t)
\sim e^{-\kappa_{\rm D}t}$). In this case the temperature is not
$\kappa/(2\pi)$ but $\kappa_{\rm eff}/(2\pi)$, with%
\begin{equation}%
\kappa_{\rm eff}:=\frac{\kappa\,\kappa_{\rm D}}{\kappa+\kappa_{\rm D}}\;,%
\label{kappaeff}%
\end{equation}%
and (at any finite time) there is neither an apparent horizon
nor an ergoregion within
the configuration.  Explanations of particle production based on
tunneling then seem not viable, and the phenomenon is closer to being
interpreted as dynamical in origin.  If fact, these configurations
interpolate between situations in which the dynamics appears more
prominently --- when $\kappa_{\rm D} \ll \kappa$ we have that the
temperature goes as $\kappa_{\rm D}/(2\pi)$ --- and others in which
the characteristics of the approached configuration are the more
relevant and ``universality'' is recovered --- when $\kappa_{\rm D}
\gg \kappa$ we have that the temperature goes as
$\kappa/(2\pi)$, indistinguishable from Hawking's result.%
\end{description}%

By looking at our simple critical model, we can say that, in
geometrical (kinematical) terms, in order to obtain a steady and
universal flux of particles from a collapsing (spherically
symmetric) star there is no need for its surface to actually cross
the Schwarzschild radius; it is sufficient that it tend towards it
asymptotically (in proper time), with sufficient rapidity.%

Our critical configurations could prove
to be relevant also in the overall picture of semiclassical collapse
and evaporation of black hole-like objects.  Our results based on
critical configurations suggest an alternative scenario to the
standard paradigm.  At this stage we are only able to present it in
qualitative and somewhat speculative terms. Being aware of the
various assumptions that could ultimately prove to be untenable, we
still think it is worth to present this possible alternative
scenario.%

Imagine a dynamically collapsing star.  The collapse process starts
to create particles dynamically before the surface of the star
crosses its Schwarzschild radius (this particle creation is normally
associated with a transient regime and has nothing to do with
Hawking's Planckian radiation). The energy extracted from the
star in this way will make (due to energy conservation) its total
mass decrease, and so also its Schwarzschild radius. By this
argument alone, we can see that a process is established in which
the surface of the star starts to closely chase its Schwarzschild
radius while both collapse towards zero (this situation was already
described by Boulware in reference~\cite{boulware}).  Now, the
question is: Will the surface of the shrinking star capture its
shrinking Schwarzschild radius in a finite amount of proper time?%

Let us rephrase this question in the language of this paper.  In an
evaporating situation our function $\xi(t)$ still represents the
distance between the surface of the star and its Schwarzschild
radius. The standard answer to the previous question is that
$\xi(t)$ becomes zero in a finite amount of proper time.  To our
knowledge, this view (while certainly plausible) is not guaranteed by
explicit systematic and compelling calculations but still relies on
somewhat qualitative arguments.  The standard reasoning can be
presented as follows: For sufficiently massive collapsing objects,
the classical behaviour of the geometry should dominate any quantum
back-reaction at any and all stages of the collapse process, as
Hawking's temperature (considered as an estimate of the strength of
this back-reaction) is very low; quantum effects would be expected
to become important only at the last stages of the evaporation
process.\footnote{In reference \cite{ashtekar-bojowald},
Ashtekar and Bojowald advocate for a different view in which they
clearly associate the quantum effects with the formation of the
singularity and not just with the last stages of the evaporation
process. In their proposed scenario the formation of a trapped
horizon does not need to imply loss of
information.}%

However, in opposition to this standard view, Stephens, 't Hooft,
and Whiting~\cite{stephens-thooft} have argued for the mutual
incompatibility of the existence of external observers measuring a
Hawking flux and, at the same time, the existence of infalling
observers describing magnitudes beyond the apparent horizon.
The reason is that the operators describing any feature of the
Hawking flux do not commute (and this non-commutation blows up at
late times) with the infalling components of the energy-momentum
tensor operator at the horizon. Therefore, if we accept this
argument, the presence of a Hawking flux at infinity would be
incompatible with the actual formation of the trapping horizon,
which would be destroyed by the back-reaction associated to Hawking
particles. This fact leads these authors (seeking for self
consistency), to look for the existence of a Hawking flux (or at
least a flux looking very much like it) in background geometries in
which the collapse process of the star is halted, just before
crossing the Schwarzschild radius, producing a bounce. (In our
language this could be represented by a function $\xi(t)$
monotonically decreasing from $t=-\infty $ to some $t=t_*$, at which
it reaches a very small positive value, and then monotonically
increasing from $t=t_*$ to $t=+\infty $.)  In their analysis they
found exactly that: An approximately Planckian spectrum of particles
present at infinity during a sufficiently long time interval.%

However, the modified behaviour that deviates the least from the
classical collapse picture, and at the same time eliminates the
trapping horizon, is that in which $\xi(t)$ does not reach zero,
but just ``asymptotically approaches zero'' at infinite proper
time, and does that very quickly.  This is represented in our
critical configurations by the exponential behaviour $\xi(t) \sim
e^{-\kappa_{\rm D} t}$ with a very large $\kappa_{\rm D}$. The
interesting point is that the analysis in this paper
suggests that with quasi-stationary configurations like this, one
could expect quasi-stationary Planckian radiation at a temperature
very close to $\kappa/(2\pi)$, just like in the Hawking
process.%

\begin{figure}[htbp]%
\vbox{ \hfil\scalebox{0.50}{{\includegraphics{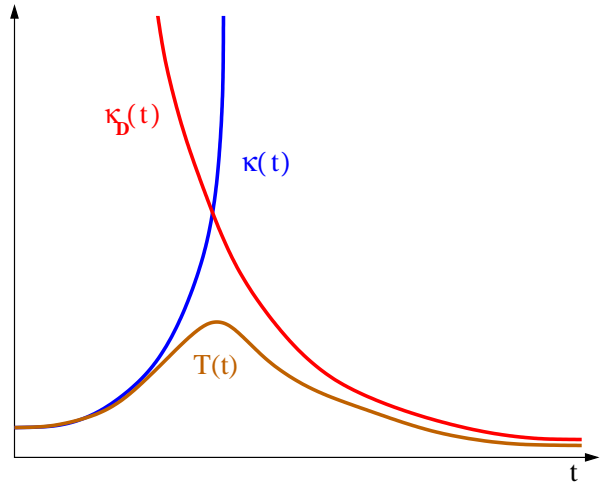}} }\hfil}%
\bigskip%
\caption{%
Possible behaviour of the evaporation temperature in an
alternative semiclassical collapse and evaporation process
based on critical configurations.%
} \label{F:critical-collapse}%
\end{figure}

Standard GR suggests that the surface gravity $\kappa$
(inversely proportional to the total mass of the star) would
increase with time through the back-reaction caused by the quantum
dissipation. Moreover, it is sensible to think that during the
evaporation process $\kappa_{\rm D}$ would also depend on $t$. As
the evaporation temperature increases ($\kappa$ increases) the
back-reaction would become more efficient and therefore we might
expect that $\kappa_{\rm D}$ decreases. Then, one could arrive at
a situation as the one portrayed in
figure~\ref{F:critical-collapse}. The evolution of the evaporation
temperature would interpolate between a starting temperature
completely controlled by $\kappa$ and a late time temperature
completely controlled by $\kappa_{\rm D}$, showing a possible
semiclassical mechanism for regularizing the end point of the
evaporation process.
In this scenario the complete semiclassical geometry will have
neither an apparent horizon nor an event horizon. In this
circumstance there would be no trans-Planckian problem, nor
information loss associated with the collapse and evaporation of
this black hole-like object. Whether this scenario is viable or not
will be the subject of future work.%

Let us end by making a brief comment concerning modified dispersion
relations. Everything we said in this paper assumes strict adherence
to Lorentz symmetry. Even if semiclassical gravity contained
Lorentz-violating traces in the form of modified dispersion
relations at high energy, one would still expect that the resulting
scenario for the collapse and evaporation of a black hole-like
object would keep the quasi-stationary Hawking-like flux of
particles as a robust prediction~\cite{Unruh}. However, the complete
conceptual scenario could be very different. In the presence of
dispersion at high energies, the notion of horizon itself shows up
only as a low-energy concept.  For example, with superluminal
modifications of the dispersion relations, high energy signals will
be able to escape from the trapped region. The non-analytic
behaviour of some sets of modes at the horizon become regularized.
Therefore, the Stephens--'t Hooft--Whiting obstruction described
above forbidding the formation of a (now approximate) trapping
horizon need no longer apply. We expect that by analyzing different
analogue models in which the Lorentz violating terms appear at
different energy scales one would be able to explore the transition
between all these alternative paradigms.%

\section*{Acknowledgements}

The authors would like to thank  V\'{\i}ctor Aldaya and Ted
Jacobson for stimulating discussions.  C.B.\ has been funded by
the spanish MEC under project FIS2005-05736-C03-01 with a partial
FEDER contribution.  C.B.\ and  S.L.\ are also supported by a
INFN-MEC collaboration.  The research of M.V.\ was funded in part
by the Marsden Fund administered by the Royal Society of New
Zealand.  M.V.\ also wishes to thank both ISAS (Trieste) and IAA
(Granada) for hospitality.%

%

{\small \begin{thebibliography}{99}%
\bibitem{analogue-book}%
M. Novello, M. Visser and G. Volovik (eds.), %
{\it Artificial Black Holes\/} %
(Singapore, World Scientific, 2002).%
\bibitem{living-review}%
C.~Barcel\'o, S.~Liberati and M.~Visser, %
``Analogue gravity,'' %
Living Rev.\ Relativity {\bf 8}, 12 (2005) %
[arXiv:gr-qc/0505065].  %
URL (cited on 22 March 2006): %
{\tt http://www.livingreviews.org/lrr-2005-12}%
\bibitem{analogue-bec}%
C.~Barcel\'o, S.~Liberati and M.~Visser, %
``Analogue gravity from Bose-Einstein condensates,'' %
Class.\ Quantum Grav.\ {\bf 18}, 1137--1156 (2001) %
[arXiv:gr-qc/0011026].%
\bibitem{analogues-of-and-for}%
M.~Visser, C.~Barcel\'o and S.~Liberati, %
``Analogue models of and for gravity,'' %
Gen.\ Relativ.\ Gravit.\  {\bf 34}, 1719--1734 (2002) %
[arXiv:gr-qc/0111111].%
\bibitem{companion}%
C.~Barcel\'o, S.~Liberati, S.~Sonego and M.~Visser, %
``Causal structure of analogue spacetimes,'' %
New J.\ Phys.\ {\bf 6}, 186 (2004) %
[arXiv:gr-qc/0408022].%
\bibitem{hawking}%
S.~W.~Hawking, %
``Black hole explosions,'' %
Nature {\bf 248}, 30--31 (1974); %
\\[2mm]%
--------- %
``Particle creation by black holes,'' %
Commun.\ Math.\ Phys.\ {\bf 43}, 199--220 (1975); %
Erratum: {\em ibid.\/} {\bf 46}, 206 (1976). %
\bibitem{bd}%
N.~D.~Birrell and P.~C.~W.~Davies, %
{\em Quantum Fields in Curved Space\/} %
(Cambridge, Cambridge University Press, 1982).%
\bibitem{visser98}%
M.~Visser, %
``Acoustic black holes: horizons, ergospheres, and Hawking
radiation,'' %
Class.\ Quantum Grav.\ {\bf 15}, 1767--1791 (1998) %
[arXiv:gr-qc/9712010].%
\bibitem{hu}%
B.~L.~Hu, %
``Hawking-Unruh thermal radiance as relativistic exponential scaling
of quantum noise,'' %
in {\em Thermal Field Theory and Applications\/}, %
edited by Y.\ X.\ Gui, F.\ C.\ Khanna and Z.\ B.\ Su %
(Singapore, World Scientific, 1996), pp.\ 249--260 %
[arXiv:gr-qc/9606073].%
\bibitem{essential}%
M.~Visser, %
``Essential and inessential features of Hawking radiation,'' %
Int.\ J.\ Mod.\ Phys.\ D {\bf 12}, 649--661 (2003) %
[arXiv:hep-th/0106111].%
\bibitem{hell}%
S.~W.~Hawking and G.~F.~R.~Ellis, %
{\em The Large Scale Structure of Space-Time\/} %
(Cambridge, Cambridge University Press, 1973).%
\bibitem{wald}%
R.~M.~Wald, %
{\em General Relativity\/} %
(Chicago, University of Chicago Press, 1984).%
\bibitem{Corley}%
S.~Corley and T.~Jacobson, %
``Black hole lasers,'' %
Phys.\ Rev.\ D {\bf 59}, 124011 (1999) %
[arXiv:hep-th/9806203].%
\bibitem{Schutzhold}%
R.~Sch\"utzhold and W.~G.~Unruh, %
``Gravity wave analogues of black holes,'' %
Phys.\ Rev.\ D {\bf 66}, 044019 (2002).%
\bibitem{lrs}%
S.~Liberati, T.~Rothman and S.~Sonego, %
``Nonthermal nature of incipient extremal black holes,'' %
Phys.\ Rev.\ D {\bf 62}, 024005 (2000) %
[arXiv:gr-qc/0002019].%
\bibitem{LRS2}
  S.~Liberati, T.~Rothman and S.~Sonego, %
  ``Extremal black holes and the limits of the third law,'' %
  Int.\ J.\ Mod.\ Phys.\ D {\bf 10}, 33--39 (2001) %
  [arXiv:gr-qc/0008018].%
\bibitem{oppenheimer-snyder}%
J.~R.~Oppenheimer and H.~Snyder, %
``On continued gravitational contraction,'' %
Phys.\ Rev.\ {\bf 56}, 455--459 (1939).%
\bibitem{pg}%
P.~Painlev\'e, %
``La m\'ecanique classique et la theorie de la relativit\'e,'' %
C.\ R.\ Acad.\ Sci.\ (Paris) {\bf 173}, 677--680 (1921).  %
\\[2mm]%
A.~Gullstrand, %
``Allgemeine L\"osung des statischen Eink\"orperproblems in der
Einsteinschen Gravitationstheorie,'' %
Ark.\ Mat.\ Astron.\ Fys.\ {\bf 16}, 1--15 (1922).%
\bibitem{parker}%
L.~Parker, %
``The production of elementary particles by strong gravitational
fields,'' %
in {\em Asymptotic Structure of Space-Time\/}, %
edited by F.\ P.\ Esposito and L.\ Witten %
(New York, Plenum, 1977), pp.\ 107--226.%
\bibitem{volovik-ripplons}%
G.~E.~Volovik, %
``Black-hole horizon and metric singularity at the brane separating
two sliding superfluids,'' %
Pisma Zh.\ Eksp.\ Teor.\ Fiz.\ {\bf 76}, 296--300 (2002) %
[JETP Lett.\ {\bf 76}, 240--244 (2002)] %
[arXiv:gr-qc/0208020].%
\bibitem{boulware}%
D.~G.~Boulware, %
``Hawking radiation and thin shells,'' %
Phys.\ Rev.\ D {\bf 13}, 2169--2187 (1976).%
\bibitem{ashtekar-bojowald}%
A.~Ashtekar and M.~Bojowald, %
``Black hole evaporation: a paradigm,'' %
Class.\ Quantum Grav.\ {\bf 22}, 3349--3362 (2005) %
[arXiv:gr-qc/0504029].%
\bibitem{stephens-thooft}%
C.~R.~Stephens, G.~'t Hooft and B.~F.~Whiting, %
``Black hole evaporation without information loss,'' %
Class.\ Quantum Grav.\  {\bf 11}, 621--647 (1994) %
[arXiv:gr-qc/9310006].%
\bibitem{Unruh}%
W.~G.~Unruh and R.~Sch\"utzhold, %
``On the universality of the Hawking effect,'' %
arXiv:gr-qc/0408009.%

\end{thebibliography}}
\end{document}